\documentclass[conference]{IEEEtran}
\IEEEoverridecommandlockouts
\usepackage{cite}
\usepackage{amsmath,amssymb,amsfonts}
\usepackage{algorithmic}
\usepackage[pdftex]{graphicx}
\usepackage{textcomp}
\usepackage{xcolor}
\def\BibTeX{{\rm B\kern-.05em{\sc i\kern-.025em b}\kern-.08em
    T\kern-.1667em\lower.7ex\hbox{E}\kern-.125emX}}

\usepackage{amsfonts} 
\usepackage{amsmath}
\usepackage{amsthm}
\usepackage{dsfont}

\usepackage{tikz} % the tikz package  
\usetikzlibrary{shapes.geometric,arrows} % to implement the corresponding shapes and arrows, these are used.  
\usetikzlibrary{positioning}

\usepackage{hyperref}

\theoremstyle{definition}

\begin{document}

\title{TangleSim: An Agent-based, Modular Simulator for DAG-based Distributed Ledger Technologies}

\author{
\IEEEauthorblockN{Bing-Yang {Lin}\IEEEauthorrefmark{1}, Daria {Dziubałtowska}\IEEEauthorrefmark{1}, Piotr {Macek} \IEEEauthorrefmark{1}, Andreas {Penzkofer}\IEEEauthorrefmark{1}, Sebastian {Müller}\IEEEauthorrefmark{2}}
\IEEEauthorblockA{\IEEEauthorrefmark{1}IOTA Foundation, Berlin, Germany, Email: research@iota.org}
\IEEEauthorblockA{\IEEEauthorrefmark{2}Aix-Marseille Universit\'e, 
 CNRS, 
 I2M, UMR 7373, 13453 Marseille, France, Email: sebastian.muller@univ-amu.fr}
}

\IEEEoverridecommandlockouts

%\IEEEpubid{\makebox[\columnwidth]{979-8-3503-1019-1/23/\$31.00~\copyright2023 IEEE \hfill} \hspace{\columnsep}\makebox[\columnwidth]{ }}

\maketitle
%\IEEEpubidadjco

\pagestyle{empty}

\begin{abstract}
 DAG-based DLTs allow for parallel, asynchronous writing access to a ledger. Consequently, the perception of the most recent blocks may differ considerably between nodes, and the underlying network properties of  the P2P layer have a direct impact on the performance of the protocol. Moreover, the stronger inter-dependencies of several core components demand a more complex and complete approach to studying such DLTs.
 
This paper presents an agent-based, open-sourced simulator for large-scale networks that implement the leaderless Tangle 2.0 consensus protocol. Its scope includes modelling the underlying peer-to-peer communication with network topology, package loss, heterogeneous latency, the gossip protocol with reliable broadcast qualities, the underlying DAG-based data structure, and the consensus protocol. 

The simulator allows us to explore the performance of the protocol in different network environments, as well as different attack scenarios. 
\end{abstract}

\IEEEpeerreviewmaketitle

\section{Introduction}
\label{sec:intro}

Distributed ledger technology (DLT) is a category of protocols in distributed systems that employs a consensus algorithm as a core component \cite{consensusProtocols}. Due to the high complexity of large-scale DLT systems, the performance evaluation of these protocols is generally challenging and costly \cite{simulatorOverview}. Moreover, blind development of a DLT system without proper performance evaluation may introduce vulnerabilities in the actual deployment phase \cite{blockchainEvaluation}. To enable the evaluation of a DLT system in the early development stage, as well as validate and test the system in different environments with different parameter settings, a DLT simulator is a necessary ingredient to design a secure protocol.
\begin{figure}[ht]
    \centering
    \includegraphics[width=0.45\textwidth]{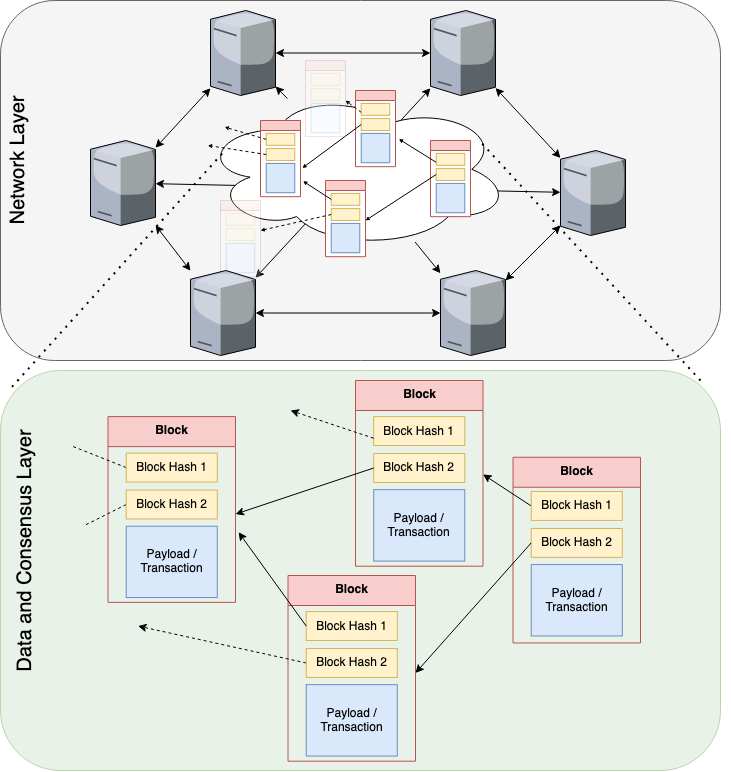}
    \caption{Overview of the tree layers in the simulator: network, data, and consensus. Since the consensus is based on the data layer, both layers are shown as one.}
    \label{fig:overview}
\end{figure}

Only recently, the development and application of blockchain simulators have received significant attention \cite{simulatorOverview}. For example, in \cite{IoTRequirement}, the opinions from various experts regarding creating a simulation environment for IoT-based blockchain applications were gathered and studied. 
The study indicates that, at present, there is a lack of available simulators to highlight different requirements and perform the necessary analysis.
A step towards a conceptual design of simulators was made in \cite{VademecumBlockchain} and \cite{IoTSurvey}  by abstracting the system into five layers, including network, consensus, data, execution, and application layers. Even with such a modular architecture, developing a generic blockchain simulator is still challenging.  Currently, no simulator fully covers the wide operational range of features and capabilities of existing blockchain technologies. Considering its complexity, developing a generic simulator for various blockchain technologies is impractical. However, developing a simulator with valuable features suitable for specific analysis and applications is practical and brings considerable benefits to designing a robust and performant protocol.

\subsection{Related Work}

The first blockchain simulator was proposed in 2013 \cite{EvaluatingBitcoin} to evaluate user privacy in Bitcoin. Later, in 2019, BlockSim \cite{BlockSim} proposed a simulation framework for various blockchain systems, which was proposed as a framework and software tool. To provide more realistic results than BlockSim, BlockPerf \cite{BlockPerf} was proposed to cover the five layers mentioned in the introduction more extensively. \cite{SegWitBlockSim} further enhanced the BlockSim by introducing Segregated Witness \cite{bip0141, bip0144}, one of the most crucial soft-fork upgrade implementations. \cite{CBlockSim} modularized the Blocksim based on five blockchain processes defined in \cite{consensusProtocols} to achieve high performance and scalability.

Recently, \cite{SimulatorsMapping} provided a systematic mapping review of simulators focusing on usability, reliability, capabilities, and supported features. 

More blockchain simulators were surveyed and reviewed in \cite{systematicReview, CBlockSim}. However, till today, only three Directed-Acyclic-Graph-based (DAG-based) simulators can be found in the literature. The first DAG-based simulator was TangleSimulator \cite{TangleSimulator} in 2018 to study the Tangle whitepaper version \cite{theTangle2016}. As the first proposed DAG-based simulator, it only supported limited configuration options. To enhance the TangleSimulator, CIDDS \cite{CIDDS} was proposed as the second DAG-based simulator in the same year to investigate the influence of the networking layer and throughput of the system. The third DAG-based simulator is DAGSim \cite{DAGsim}. It is a continuous-time multi-agent simulation framework for DAG-based cryptocurrencies. However,  all the nodes are simulated in serial, hence, not allowing the simulation of the most critical aspects of the asynchronous nature. 

The above DAG-based simulators analyze the behaviours of the outdated IOTA protocol \cite{theTangle2016}, and none implemented and supported conflicts analysis or any adverse scenario. In addition, only a few metrics, including transaction attachment probabilities and basic statistics, have been presented in the simulators so far. The block confirmation time, one of the most critical metrics, was not addressed yet.
%They also do not support and implement the up-to-date IOTA consensus protocol proposed in 2022 \cite{muller2022}.

\subsection{Contribution}\label{sec:contribution}
We propose TangleSim, the first DAG-based simulator that analyzes the network, data, and consensus layers and their interplay. 
TangleSim supports the latest IOTA consensus protocol \cite{muller2022}, along with various performance metrics, to investigate the details of the DAG-based protocol behaviours.
While simulations in multi-core server environments are a typical approach, we consider the accessibility of the simulator a vital design criterion.  For this reason, we propose a deceleration mechanism to allow large-scale simulations on one local machine without using high-performance hardware.

\section{Fundamentals of the protocol}\label{sec: consensus}

In a blockchain, each block contains a record of multiple transactions, and each block is cryptographically linked to the previous block in the chain using a cryptographic hash function. This creates a chain of blocks that cannot be altered without also altering all subsequent blocks in the chain, requiring the consensus of the network. The chain of blocks is constantly growing as new transactions are added to the network, creating an immutable record of all transactions that have occurred on the blockchain. 

In a DAG-based DLT, blocks can be linked to more than one previous block creating a DAG of blocks instead of a chain of blocks; see also Figure \ref{fig:overview}. The data structure consists of blocks and references from more recent blocks, called \textit{children}, to older blocks, called \textit{parents}. The arising block-DAG is also called \textit{Tangle}. 
% As shown in Figure \ref{fig:Tangle}, the genesis block is the oldest block in the Tangle, which has no parents. 
\textit{Tips} are blocks without any references to them, i.e., they are yet without children. The Tangle starts with a genesis block, and nodes attach new blocks to previous blocks according to an algorithm called \textit{tip selection}. %The maximum number of tips (or parents) references is limited to $n$. %In Figure \ref{fig:Tangle}, the parent count is defined as $=2$.  Although blocks can generally contain more than one transaction, we assume that each block contains a single transaction and therefore identify blocks with the transactions they contain. 

As in a blockchain, these references to previous parts of the DAG allow for defining a certain level of confidence for the inclusion of the blocks into the final data structure. In the example of the Tangle 2.0 consensus protocol,  \cite{muller2022}, the level of endorsement of blocks and their associated transactions are defined by the  \textit{Approval Weight}. More precisely, the Approval Weight of a block (or the transactions it contains) is calculated as the cumulative weight of all issuers of blocks referencing (directly or indirectly)  the given block. %These weights are normalized so that the sum of the total weights is $1$. 
This is a natural extension of the longest chain rule in blockchains or the cumulative weights in the original proposal of the Tangle~\cite{theTangle2016}. A block is confirmed when its cumulative weight reaches a threshold $\theta$.

\section{Simulation Setup and Definitions}\label{sec: simulation setup}

\subsection{Data Flow of the Simulator}

\begin{figure}[!ht]
    \centering
    \includegraphics[width=0.45\textwidth]{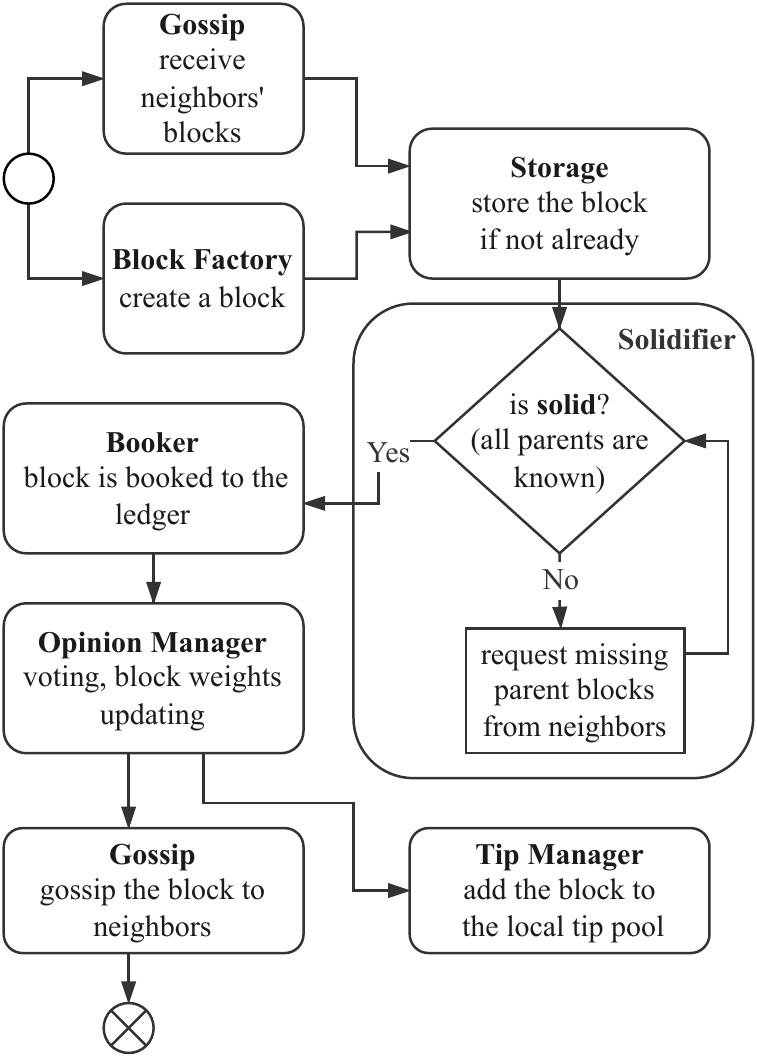}
    \caption{Block flow diagram in the simulator, from start ($\bigcirc$) to end ($\bigotimes$) of a block's processing life cycle.}
    \label{fig:blockflow-simulator}
\end{figure}

The simulator's design focuses on the essential components relevant to the prominent performance measures. Specifically, we implement the necessary components to support basic network features, 
such as gossiping and scheduling and requesting missing blocks, the DAG-based data structure properties, and the consensus layer.

In the following, we describe the implemented components and refer to  Figure~\ref{fig:blockflow-simulator} for the interplay of some components.

\textbf{Block Factory.}
The block factory creates a valid block. It uses a tip selection algorithm to select valid tips. In the full node implementation, it is also responsible for the block signing and, if applied, performing Proof-of-Work. In the simulator, these last two operations are omitted.  We note that these two operations lead to a larger delay between choosing the references and the time the other participants see the block, influencing the protocol's performance.  However, we can model this effect by increasing the network delay accordingly.

\textbf{Storage.}
The storage module stores blocks in a database on a disk. In the simulator, all blocks are stored in memory, which improves performance but limits the size of the block DAG that can be stored.

\textbf{Solidifier.}
\emph{Solidification} is the process of requesting missing blocks. A missing block is identified by comparing whether the blocks to which a received block references are present in the node's database. If any parent block is missing, the module sends a request to its neighbours asking for that missing block.  
Once all directly or indirectly referenced blocks are present in the database for a given block, the block is marked as \textit{solid}.

\textbf{Booker.}
The Booker is a module responsible for adding blocks into the Tangle. It identifies and manages double spendings.

\textbf{Opinion Manager.}
The Opinion Manager indicates which conflicts are currently supported by the node. It does not need to constantly keep track of all the conflicts and their Approval Weights but instead is lazily evaluated when a node needs to issue a vote. In the simulator, the logic is greatly simplified due to the fact that it supports only one conflict set.

\textbf{Gossip.}
The gossip module is responsible for disseminating new blocks between the nodes in the underlying P2P network.

\textbf{Tip Manager.}
The Tip Manager is a module responsible for keeping track of available valid tips and returning tips to be used in the block factory module for newly created blocks.

\begin{figure*}[t!]
    \centering
    \includegraphics[width=0.8\textwidth]{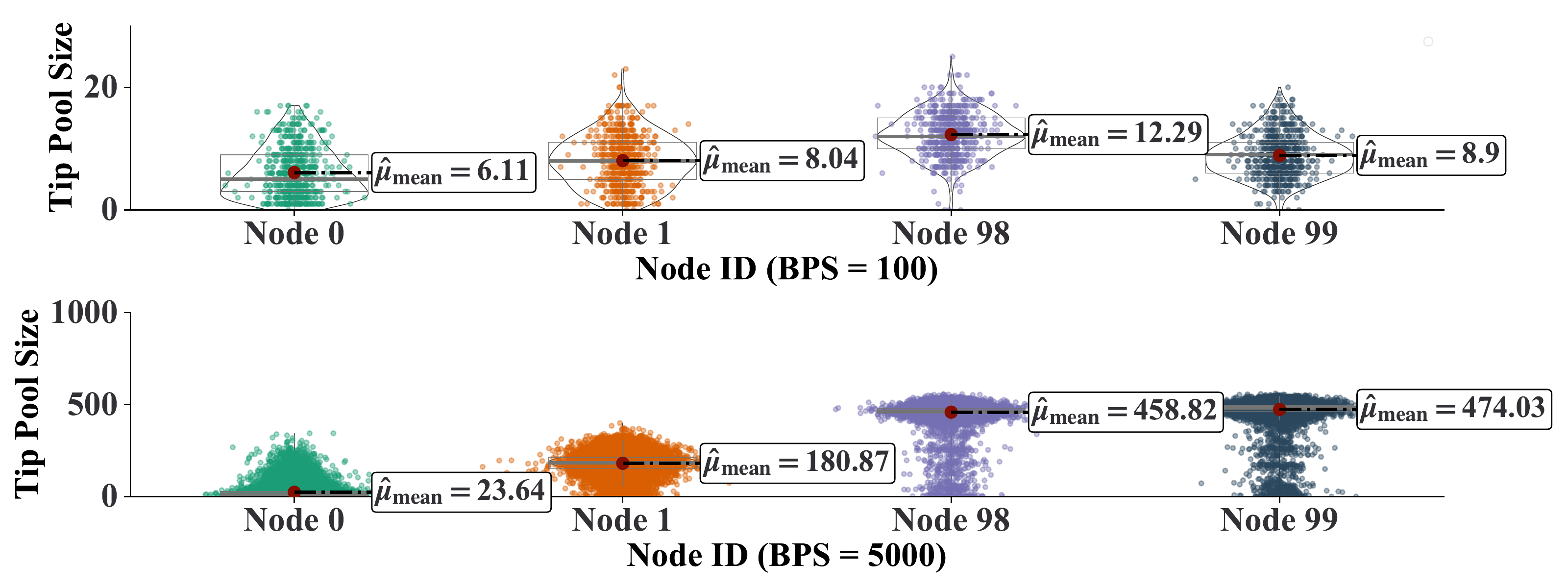}
    \caption{Different perception of the local tip pools for two different values of BPS.}
    \label{fig:TP}
\end{figure*}

\section{Design and Features}\label{sec: experiment}
We include the three layers, networking, data, and consensus, in the simulator design. We briefly describe the implemented features for each of these layers and refer to \cite{otv-simulator} for a more detailed presentation.

\subsection{Network Layer}
The node count $N$ is the number of nodes in the simulation. Every node or peer carries a certain weight that serves as a Sybil protection, controls the writing access, and is used for the consensus finding. The simulator allows the modelling of the weight of the nodes with a Zipf law with the parameter $s$.  For example, a homogeneous network can be modelled using $s=0$, while for $s$ significantly above $1$, a more centralized weight distribution can be investigated. The block issuance rate of a node is, by default, assumed to be proportional to its weight but can be chosen differently as well. 
\begin{table}[htbp]
    \caption{Parameters of The Network Layer}
    \begin{center}
    \begin{tabular}{|c|l|c|}
         \textbf{Symbol} & \textbf{Description} & \textbf{Default Value} \\
         $N$ & Node count & 1,000 \\
         $s$ & Zipf's parameter of weight distribution & 0.9 \\
         $\gamma$ & Watts-Strogatz rewiring probability & 100\% \\
         $k$ & Number of neighbors & 8 \\
         $d\textsubscript{min}$ & Minimum delay & 50ms \\ 
         $d\textsubscript{max}$ & Maximum delay & 150ms \\ 
         $p\textsubscript{loss}$ & Packet loss & 0 \\
        %  $n_q$ & A list of number of adversary nodes & [5, 5] \\
         \end{tabular}
    \label{tab:NetworkLayer}
    \end{center}
\end{table}
We model the topology of the P2P network using Watz-Strogatz graphs \cite{watts1998collective}.  This class of graphs allows for a high level of controllability of parameters, such as the number of each node's neighbours, the network randomness, and the network diameter. To mimic real-world behaviour, the simulator specifies packets' delay between neighbouring peers and packet loss for each node's connection. The blocks are distributed via a gossip algorithm whose rate is controlled by a scheduler module. In addition, we implement a soldifier module that requires missing blocks from the nodes' neighbours.  

A more subtle influence on the protocol performance, as demonstrated in \cite{cullen2019-variable-delay}, is the randomness of the delay. The delay between the P2P communication is modelled by a uniform random variable in the interval $[d\textsubscript{min}, d\textsubscript{max}]$ and each package can be lost with a probability $p\textsubscript{loss}$. We refer to Table \ref{tab:NetworkLayer} for an overview of the different parameters.

\subsection{Data Layer}
The data layer and, in particular, the structure of the DAG depends on the number  of tips $n_{tips}$ every block refers to. Another essential parameter is the system's throughput that is controlled by the blocks per second $\mathrm{BPS}$. 
In addition, we assume the block issuance time interval of nodes follows a Poisson distribution \cite{Penzkofer2021ImpactOD}. We refer to Table \ref{tab:DataLayer} for an overview of these variables.

\begin{table}[htbp]
    \caption{Parameters of the data layer}
    \begin{center}
    \begin{tabular}{|c|l|c|}
         \textbf{Symbol} & \textbf{Description} & \textbf{Default Value} \\
         $n\textsubscript{tips}$ & Tips count & 8 \\
         $\textrm{BPS}$ & Blocks per second  & 100 \\
         $\textrm{IMIF}$ & Distribution of delay between blocks & ``Poisson'' \\
        %  $n_q$ & A list of number of adversary nodes & [5, 5] \\
         \end{tabular}
    \label{tab:DataLayer}
    \end{center}
\end{table}

\subsection{Consensus Layer}

% Introduce default values, what the meaning of each parameter is and how it impacts the results
% Introduce the plan of experiments in more detail.

The confirmation threshold $\theta$ controls the confirmation of the blocks; a block will be confirmed if it gets an Approval Weight larger than $\theta$. The simulator allows the modelling of various Byzantine environments. The adversarial weight $q_a$ describes the proportion in control of an adversary, and $d_q$ is the delay of the adversarial messages.  This allows us to model attackers that impact the P2P layer. See Table \ref{tab:ConsensusLayer} for an overview.

\begin{table}[htbp]
    \caption{Parameters of the consensus layer}
    \begin{center}
    \begin{tabular}{|c|l|c|}
         \textbf{Symbol} & \textbf{Description} & \textbf{Default Value} \\
         $\theta$ & Confirmation threshold & 66\% \\
         %$\theta_c$ & confirmation threshold criterion & absolute \\
         %$d\textsubscript{ds}$ & Double spending delay & 20s \\ 
         $d_q$ & Delay from adversary & 100 ms \\
         $q_a$ & Adversary weight  & 5\% \\
         $N_a$ & Node count of adversary & 2 \\
        %  $n_q$ & A list of number of adversary nodes & [5, 5] \\
         \end{tabular}
    \label{tab:ConsensusLayer}
    \end{center}
\end{table}

We implement a particularly efficient adversary strategy, as pointed out in \cite{muller2022}. In this so-called bait-and-switch attack, the attacker issues an endless stream of conflicting transactions to prevent the ``honest weight'' from accumulating on one common transaction.

\section{Performance Measures}

\subsection{Confirmation Time}
\begin{figure}[h!]
    \centering
    \includegraphics[width=0.45\textwidth]{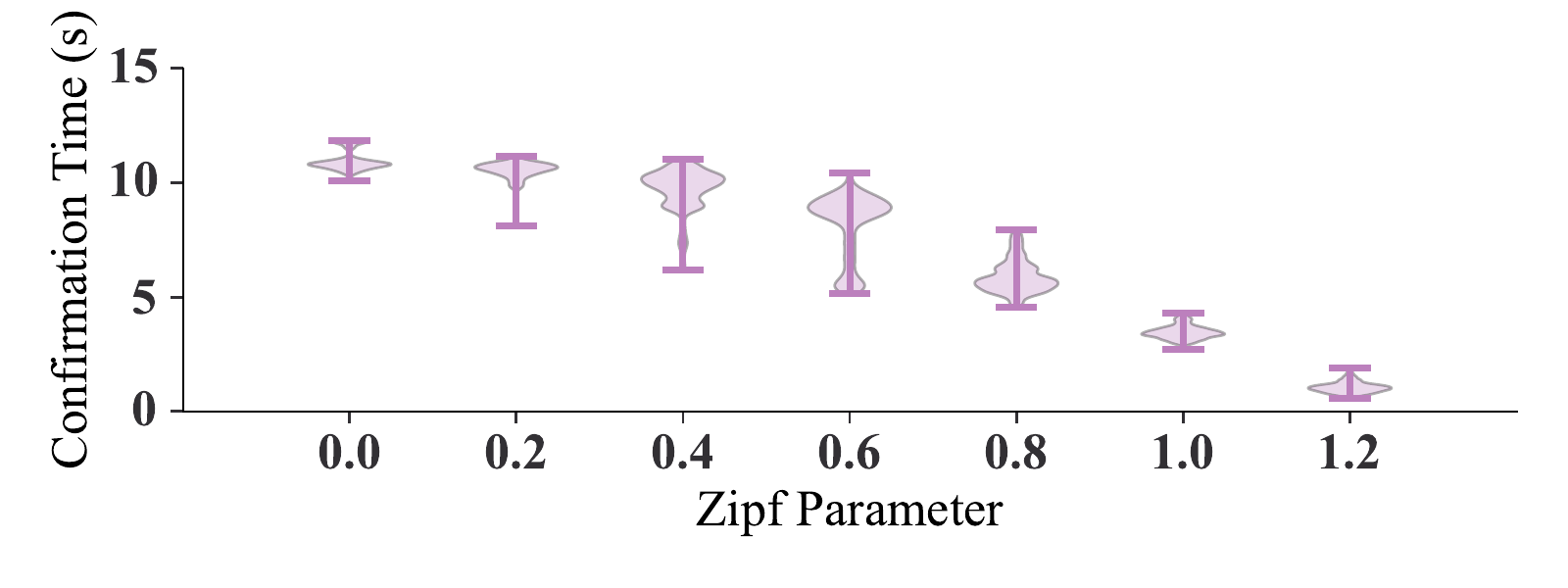}
    \caption{Dependence of the confirmation time on the weight distribution of the nodes. }
    \label{fig:CT1}
\end{figure}

The confirmation time of transactions is probably the most prominent performance measure in a DLT. In our setting, it is defined as the time period from the issuance time of a block to the time when the accumulated Approval Weight of the block is greater than the confirmation threshold $\theta$. By analyzing the confirmation time in different scenarios, the optimal configuration of a Tangle, i.e.,~the set of parameters that achieve the shortest confirmation time, can be identified.  For example, in Figure \ref{fig:CT1} we study the dependence of the confirmation time on the centrality of the weight distribution while the other parameters are set to their default values. For each parameter set $100$ simulation runs are performed.

\subsection{Tip Pool Size}
A performance measure that is of particular interest for DAG-based DLTs is the number of tips. The difference in local perceptions of these tip pools  can lead to performance degradations or syncing problems. As an example, we study in Figure \ref{fig:TP} the dependence of the tip pools with two different values of BPS and $100$ nodes; all other parameters are set to default, and $100$ simulation runs. We refer to \cite{Stability} for a more detailed analysis of the local perceptions of the tip pools.

\subsection{Consensus Time}
\begin{figure}[ht]
    \centering
    \includegraphics[width=0.45\textwidth]{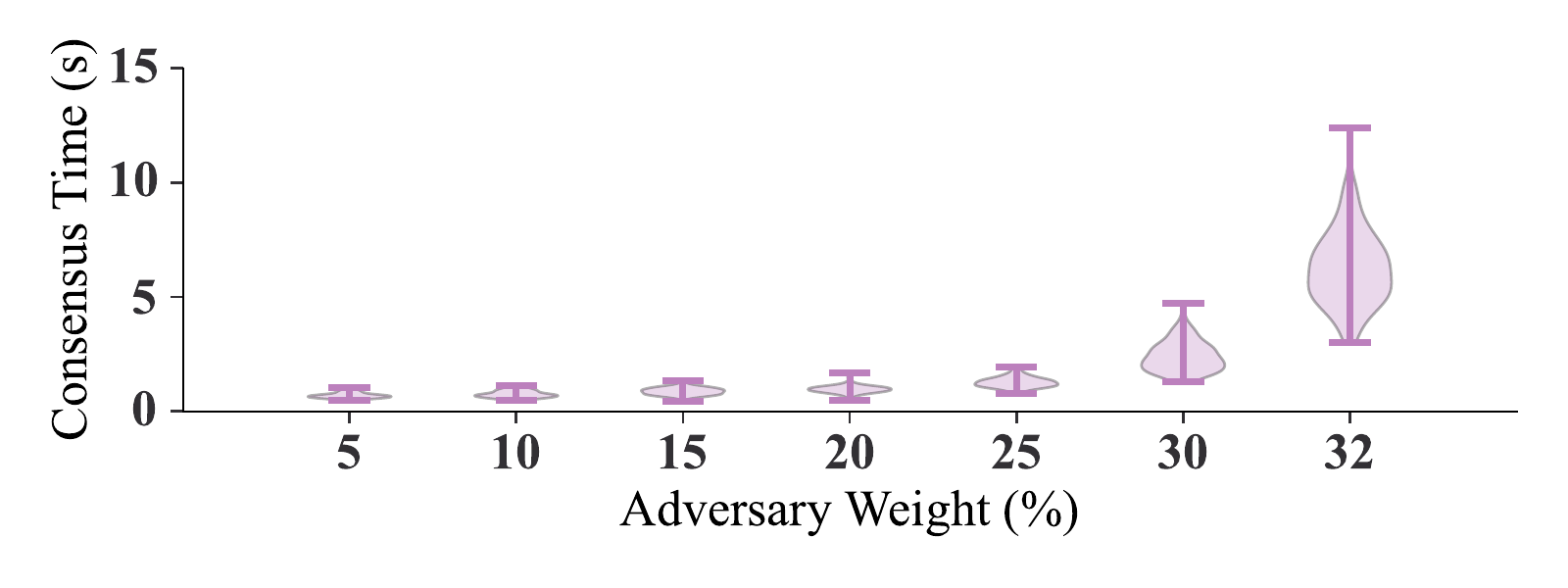}
    \caption{Dependence of the consensus time on the adversarial weight. }
    \label{fig:CS}
\end{figure}

In the case of conflicts, we are interested in the time it takes to resolve a conflict and denote it as consensus time.  As an example, we give the consensus time as a function of the adversarial weight under a bait-and-switch attack in Figure \ref{fig:CS}; the Zipf parameter is set to $2$, while the other parameters are set to default.  The study of the consensus time in different adversarial situations is complex, and we refer to \cite{robustnessTangle2.0} for a first study on the robustness of the Tangle 2.0 consensus using our simulator.

\section{Summary and Discussion}
We proposed an agent-based simulator allowing us to understand the performance of the IOTA 2.0 protocol under various network situations and evaluate the consensus protocol in a Byzantine environment.  We also presented examples of the most prominent performance measures. 
The simulator's modularity allows including additional components and attack strategies seamlessly. Moreover, it can be easily adapted to study other DAG-based DLTS and may serve as a general framework for comparing different DAG-based DLT solutions. 
\iffalse
\subsection{Slowdown factor of simulator}

The simulator is able to simulate thousands of nodes, where each node runs concurrently and interacts with its neighbours actively. However, simulating a large number of nodes exchanging blocks can overwhelm any CPU, and in order to solve this, the simulator has an additional option that allows setting a slowdown factor $ L $. This means that to simulate 1 minute of the network activity, the simulator needs $ L $ minutes. This setting needs to be adjusted for each set of simulation parameters and hardware individually, such that the hardware is maximally utilized, but not over-utilized. This makes the results reproducible even on a personal/local machine, without the need to use powerful servers. 
\fi

%\section{Conclusion}

% use section* for acknowledgment
\ifCLASSOPTIONcompsoc
   %The Computer Society usually uses the plural form
  \section*{Acknowledgments}
\else
   %regular IEEE prefers the singular form
  \section*{Acknowledgment}
\fi

The authors would like to thank Hans Moog for writing the first version of the simulator.

\bibliographystyle{IEEEtran}
\bibliography{bibliography}
\end{document}